\documentclass[12pt]{article}
\usepackage{graphicx}
\input epsf.tex
\def\DESepsf(#1 width #2){\epsfxsize=#2 \epsfbox{#1}}
\def\NPB{{ Nucl. Phys.} B}

\def\PLB{{ Phys. Lett.}  B}
\def\PRL{ Phys. Rev. Lett.}
\def\PRD{{ Phys. Rev.} D}

\def\ZPC{{ Z. Phys.} C}
\def\AP{ Astropart. Phys.}

\newsavebox{\sboxpubnumber}
\newsavebox{\sboxpubdate}
\newcommand{\pubdate}[1]{\begin{lrbox}{\sboxpubdate}{#1}\end{lrbox}}
\newcommand{\pubnumber}[1]{\begin{lrbox}{\sboxpubnumber}{\begin{tabular}{l} #1 \\
				 \usebox{\sboxpubdate}
				 \end{tabular}}
                           \end{lrbox}
                           \pubblock}
\newcommand{\Title}[1]{\begin{center} {\Large #1 } \end{center}}
\newcommand{\Author}[1]{\begin{center}{ \sc #1} \end{center}}
\newcommand{\Address}[1]{\begin{center}{ \it #1} \end{center}}

\newcommand{\pubblock}{\rightline{
			\usebox{\sboxpubnumber}}}
\newenvironment{Abstract}{\begin{quotation}  }{\end{quotation}}
\newenvironment{Presented}{\begin{quotation} \begin{center}
             PRESENTED AT\end{center}\bigskip
      \begin{center}\begin{large}}{\end{large}\end{center}
      \end{quotation}}

\begin{document}

\begin{titlepage}
\pubdate{\today}                    
\pubnumber{} 

\vfill
\Title{Supersymmetry 
and Dark Matter}
\vfill
\Author{R. Arnowitt and B. Dutta}
\Address{Center For Theoretical Physics, Department of Physics, \\
Texas A$\&$M University, College Station TX 77843-4242}
\vfill
\vfill
\vfill
\begin{Abstract}We examine supergravity models with grand unification at $M_G$ possessing R 
parity invariance. Current data has begun to significantly constrain the 
parameter space. Thus for mSUGRA, accelerator data places a lower bound on 
$m_{1/2}$ of $m_{1/2} \stackrel{>}{\sim}300$ GeV while astronomical data on the amount of relic dark 
matter narrowly determines $m_0$ in terms of $m_{1/2}$ (for fixed value of 
$\tan\beta$ 
and $A_0$) due to co-annihilation effects. Additional new data could fix the 
parameters further. Thus the parameter space is sensitive to the muon 
magnetic moment anomaly, $\delta a_\mu$, and if $\delta a_\mu$ lies 1$\sigma$ above 
its current central value,  it would exclude mSUGRA, while if it lies 
1$\sigma$ below (but is still positive) it pushes the SUSY spectrum into the 
TeV domain. The $B_s \rightarrow\mu^+ \mu^-$ decay is seen to be accessible to the Tevatron 
RunII with branching ratio sensitivity of $Br[B_s \rightarrow\mu^+ \mu^-] > 
6.5\times10^{-9}$  with $15 \rm fb^{-1}$/detector, and a value of 
$7(14)\times10^{-8}$ obtainable with $2\rm fb^{-1}$ would be sufficient to exclude mSUGRA for 
$\tan\beta < 50(55)$. Measurements of $B_s\rightarrow\mu^+ \mu^-$ can cover the full mSUGRA parameter 
space for $\tan\beta > 40$ if $\delta a_\mu > 11\times10^{-10}$, and combined 
measurements of $B_s \rightarrow\mu^+ \mu^-$, $a_\mu$ and $m_h$ (or alternately the gluino mass)
would effectively determine the mSUGRA parameters for $\mu > 0$. Detector 
cross sections are then within the range of planned future dark matter 
experiments. Non-universal models are also discussed, and it is seen that 
detector cross sections there can be much larger, and can be in the DAMA 
data region.  
\end{Abstract}
\vfill
\begin{Presented}
    DARK 2002\\
    Cape Town, South Africa \\
    4-9 February, 2002 
\end{Presented}
\vfill
\end{titlepage}
\def\thefootnote{\fnsymbol{footnote}}
\setcounter{footnote}{0}

\section{Introduction}
It is generally expected that the Standard Model will break down at 
energies above LEP, and signals of new physics will occur for energies 
$\stackrel{>}{\sim}$
100 GeV - 1 TeV. The nature of this new physics is one of the crucial 
question of particle physics. Simultaneously, astronomical data has 
determined with good accuracy the amount of dark matter in the universe, 
though the nature of that dark matter remains one of the crucial questions 
of astronomy. Supersymmetric theories with R parity invariance offer an 
explanation to both puzzles as well as a window on the cosmology of the 
very early universe at times $t\approx 10^{-7}$ sec.

Unfortunately, supersymmetric models depend upon a large number of 
parameters, and even the simplest model, mSUGRA, depends on four parameters 
and one sign. But fortunately, supersymmetry applies to a wide number of 
phenomena, and it is now becoming possible to significantly restrict the 
parameter space. The general MSSM, with over 100 free parameters (63 real 
parameters) is not very predictive. We consider here therefore, models,  
based on supergravity grand unification at $M_G \simeq2\times10^{16}$ GeV (which 
have both theoretical and experimental motivation). We examine first the 
current status of the simplest model, mSUGRA \cite{sugra1,sugra2}, and what might be 
obtained from future measurements of the muon magnetic moment, $g_\mu - 2$, 
the Higgs mass, and the B decay $B_s \rightarrow\mu^-\mu^+$. We will then look at 
non-universal models with non-universality in the Higgs or third generation 
of squarks and sleptons. For all these cases, the lightest neutralino, 
${\tilde\chi^0_1}$, is the dark matter candidate, and this will also strongly constrain 
the SUGRA parameter space.

\section{mSUGRA Model}

We briefly review the mSUGRA model which depends on four parameters and one 
sign, and thus is the most predictive of the SUSY models. We take these 
parameters to be the following: $m_0$ (the universal scalar soft breaking 
mass at $M_G$); $m_{1/2}$ (the universal gaugino mass at $M_G$); $A_0$ (the universal 
cubic soft breaking mass at $M_G$); and $\tan\beta= <H_2> / <H_1>$ (the ratio of 
the two Higgs VEVs at the electroweak scale). The sign of the Higgs mixing 
parameter $\mu$ (appearing in the superpotential as $\mu H_1 H_2$ ) is the 
remaining parameter.  (We note at the electroweak scale that the $\tilde\chi^0_1$ and 
lightest chargino, $\tilde\chi^\pm_1$, masses are related to $m_{1/2}$ by 
$m_{\tilde\chi^0_1}\stackrel{\sim}{=}0.4m_{1/2}$ and 
$m_{\tilde\chi^\pm_1}\stackrel{\sim}{=} 0.8 m_{1/2}$.) We examine this model with the following parameter 
ranges: $m_0 \leq 1$ TeV; $m_1/2 \leq 1$ TeV; $2\leq \tan\beta \leq55$; and
$A_0\leq 4m_{1/2}$. 
The above bound on $m_{1/2}$ corresponds to a gluino mass range of 
$m_{\tilde g}  \leq 
2.5$ TeV, which is also the upper mass reach for the LHC.

In the early universe, the neutralino can annihilate via s-channel $Z$, 
and  $h$, $H$, and $A$ neutral Higgs bosons  ($h$  is the light Higgs, and $H (A)$ 
are the heavy CP even (odd) Higgs), and also through t-channel sfermion 
diagrams. However, if a second particle becomes nearly degenerate with the 
$\tilde\chi^0_1$, one must include it in the early universe annihilation processes. This 
leads to the co-annihilation phenomena. In SUGRA models with gaugino grand 
unification, this accidental near degeneracy occurs naturally for the light 
stau, $\tilde \tau_1$. One can see this analytically for low and intermediate 
$\tan\beta$, where the renormalization group equations (RGE) can be solved 
analytically. One finds for $\tilde e_R$, the right selectron and the 
$\tilde\chi^0_1$ at the electroweak scale the 
results:
\begin{eqnarray}
 m_{\tilde e_R}&=&m_0^2 + 0.15 m_{1/2}^2  - sin^2\theta_W M_W^2
 \cos2\beta\\\nonumber m_{\tilde\chi^0_1}&=& 0.16m_{1/2}^2 
\end{eqnarray}
where the last term of Eq(1) is approximately (37GeV)$^2$. Thus for $m_0 = 0$, 
$\tilde e_R$ becomes degenrerate with $\tilde\chi^0_1$ at $m_{1/2} 
\stackrel{\sim}{=}$370 GeV, and co-annihilation 
thus begins at $m_{1/2} \stackrel{\sim}{=}$(350GeV - 400GeV). As $m_{1/2}$ increases, 
$m_0$ must be 
raised in lock step (to keep $m_{\tilde e_R} > m_{\tilde\chi^0_1}$). More precisely, it is the 
$\tilde\tau_1$ 
which is the lightest slepton and this particle dominates the 
co-annihilation phenomena. In general, co-annihilation implies that one 
ends up with relatively narrow allowed corridors in the $m_0 - m_{1/2}$ plane 
with $m_0$ closely correlated with $m_{1/2}$, increasing as $m_{1/2}$ increases.

Dark matter detection of Milky Way neutralinos incident on the Earth 
depends upon the neutralino - proton cross section.  For detectors with 
nuclear targets containing heavy nuclei, this is dominated by the spin 
independent cross section. The basic quark diagrams involve s-channel 
squarks, and $t$-channel $h$ and $H$ diagrams. Thus 
$\sigma_{\tilde\chi^0_1-p}$ decreases with 
increasing $m_{1/2}$ and $m_0$ (which as we have seen above increase together), and 
also increases with $\tan\beta$ (due to the Higgs couplings to the $d$ quark). 
Thus the maximum cross section will occur at high $\tan\beta$, and low 
$m_{1/2}$, $m_0$.

In order to carry out calculations in SUGRA models accurately, it is 
necessary to take into account a number of corrections, and we list the 
important ones here:

We use two loop gauge and one loop Yukawa RGE from $M_G$ to the electroweak 
scale (which we take as $(\tilde t_1 \tilde t_2)^{1/2} )$, and QCD RGE below for 
the light 
quark contributions.
Two loop and pole mass corrections are included in calculation of $m_h$.
One loop corrections to $m_b$ and $m_\tau$ \cite{rattazi,carena} are included, which are 
important for large $\tan\beta$.
Large $\tan\beta$ NLO SUSY corrections to $b \rightarrow s\gamma$ 
\cite{degrassi,carena2} are included.
All stau-neutralino co-annihilation channels are included in the relic 
density calculation with analysis valid for the large $\tan\beta$ regime 
\cite{bdutta,ellis,gomez}.

Note that we do not include Yukawa unification or proton decay constraints 
in the analysis as these depend sensitively on post-GUT physics, about 
which little is known. In fact in string or M-theory analyses with grand 
unification, while the unification of the coupling constants occur as in 
SUGRA models, the Yukawa unification or proton decay constraints do not 
generally hold \cite{green}.

\begin{figure}[htb]
\centerline{ \DESepsf(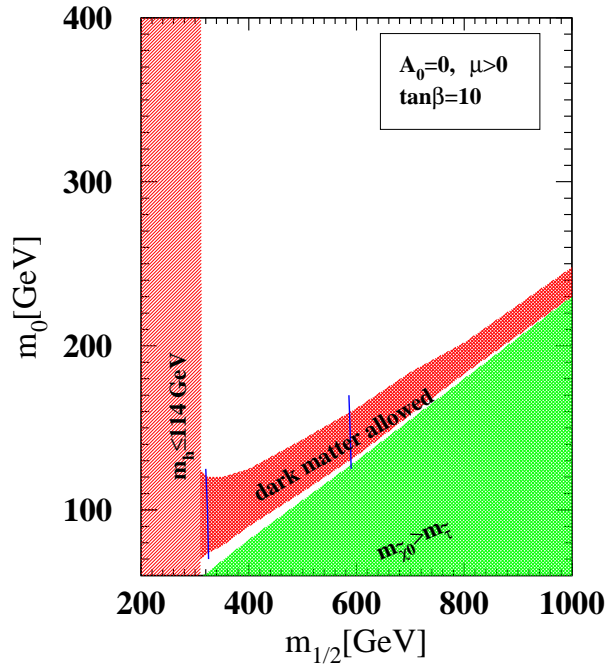 width 7 cm) }
\caption {\label{fig1}  $m_0 - m_{1/2}$ plane showing the allowed parameter region for $\tan\beta =$ 
10, $A_0 = $0, $\mu >0$. The short vertical lines give 
$\sigma_{\tilde\chi^0_1-p}=  5\times 10^{-9}$ pb 
(left line) and $1\times10^{-9}$ pb (right line).}
\end{figure}

 \begin{figure}[htb]
\centerline{ \DESepsf(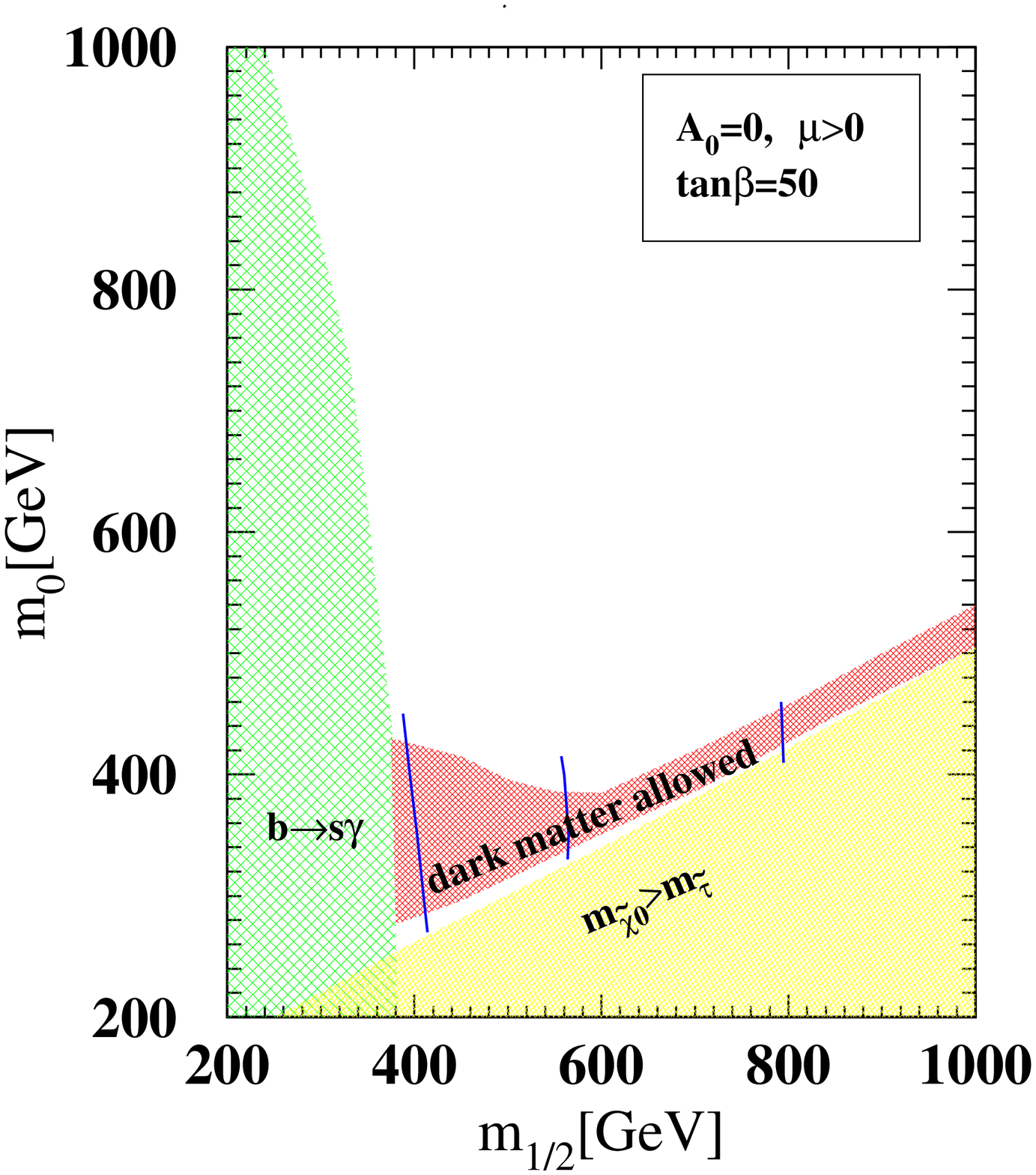 width 7 cm) }
\caption {\label{fig2}  $m_0 - m_{1/2}$ plane showing the allowed parameter region for $\tan\beta =$ 
50, $A_0 = 0$, $\mu > 0$.The short vertical lines give (from left to right) 
$\sigma_{\tilde\chi^0_1-p}= 5\times10^{-8}$ pb, $1\times10^{-8}$ pb,
$2\times10^{-9}$ pb.}
\end{figure}
\section{Current Experimental Constraints}

In order to see what the currently allowed parameter space is, one must 
impose all the present experimental constraints. However, three of these 
acting together produce significant limitations, and we mention these here:

(1) Higgs mass.
The current LEP bound on the light Higgs is $m_h >$ 113.5 GeV \cite{higgs1}. However, 
the theoretical calculation of $m_h$ \cite{higgs} may have a 2 -3 GeV error, and so we 
will conservatively interpret this bound to mean $m_h$(theory) $> 111$ GeV.

(2) $b\rightarrow s\gamma$ decay.
There is some model dependence in extracting the $b\rightarrow s\gamma$ branching 
ratio from the CLEO data, and so we will take a relatively broad range 
around the current CLEO  central value \cite{bsgamma}:
\begin{equation}
                             1.8\times10^{-4} < Br (B\rightarrow X_s\gamma) < 
4.5\times10^{-4} 
\end{equation}

(3) $\tilde\chi^0_1$ relic density.
The relic density is measured in terms of $\Omega = \rho/ \rho_c$ where $\rho$ is 
the mass density, $\rho_c = 3H^2/8\pi G_N$ and $H = (100km/sMpc)h$ is the Hubble 
constant. Analyses of the cosmic microwave background now gives a fairly 
accurate measurement of the amount of CDM, i. e. $\Omega_{\rm CDM} h^2= 
0.139\pm0.026$ \cite{turner}. We take a $2\sigma$ range around the central value:
\begin{equation}0.07 < \Omega_{\tilde\chi^0_1} h^2 < 0.21 
\end{equation}

These three constraints now combine to greatly restrict the mSUGRA 
parameter space. Thus the $m_h$ bound for low $\tan\beta$ and the $b\rightarrow s\gamma$ 
constraint for higher $\tan\beta$ produce a lower bound on $m_{1/2}$ over the entire 
parameter space of
\begin{equation}
                             m_{1/2}\stackrel{>}{\sim}(300 - 
400)\,{\rm GeV} 
\end{equation}
and consequently $m_{\tilde\chi^0_1}\stackrel{>}{\sim} (120 - 160)$ GeV. This means that most of the 
parameter space is in the $\tilde\tau_1 - \tilde\chi^0_1$ co-annihilation domain in the relic 
density calculation, and thus to satisfy the relic density bound , $m_0$ is 
approximately determined by $m_{1/2}$ (for fixed $\tan\beta$, $A_0$). This implies 
that as $m_{1/2}$ increases, so does $m_0$, and so generally on has that 
$\sigma_{\tilde\chi^0_1-p}$ 
is a deceasing function of $m_{1/2}$.

We consider first $\mu > 0$. Figs. 1 and 2 exhibit the effects discussed 
above. Thus in Fig. 1 for $\tan\beta = 10$, $A_0 =0$, one sees that the Higgs mass 
bound requires $m_{1/2} \stackrel{>}{\sim}300$ GeV, and one sees the narrow 
$m_0$ band allowed by 
the co-annihilation effects. The short vertical lines show the expected 
dark matter detection cross sections of $\sigma_{\tilde\chi^0_1-p}= 
5\time 10^{-9}$ pb (left) and 
$1\times 10^{-9}$ pb (right). Thus $m_0$ is determined by $m_{1/2}$ to within $\sim 40$GeV. Fig. 2 
shows the corresponding situation for $\tan\beta = $50. Here the 
$b\rightarrow s\gamma$  produces the low energy cut off of 
$m_{1/2} \stackrel{>}{\sim}350$ GeV. There is also 
a bulge in the dark matter allowed domain at low $m_{1/2}$ where co-annihilation 
has not yet set in. The cross sections are (left to right) $5\times10^{-8}$ pb, 
$1\times10^{-8}$ pb and $2\times10^{-9}$ pb. These cross sections are within range of future 
planned detector experiments.

\begin{figure}[htb]
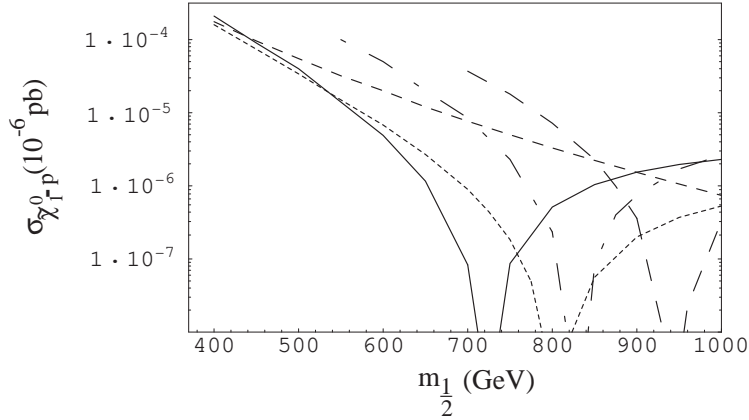

\centerline{ \DESepsf(aadcoan61020.epsf width 10 cm) }
\caption {\label{fig3} Minimum values of $\sigma_{\tilde\chi^0_1-p}$ as a function of $m_{1/2}$ for $\mu < 0$, 
$\tan\beta = 6$ (dashed), 8 (dotted), 10 (solid) and 25 (large dash)[7].}
\end{figure}

For $\mu < 0$ an accidental cancellation in $\sigma_{\tilde\chi^0_1-p}$
 occurs over a wide range 
of $\tan\beta$ \cite{ellis3,bdutta} giving large regions where $\sigma_{\tilde\chi^0_1-p}
 < 10^{-10}$ pb and 
hence probably inaccessible to future detectors. This is exhibited in Fig. 
3 \cite{bdutta}, where the minimum cross sections are plotted as one scans the 
allowed parameter space, for $\tan\beta = $6 (dashed), 8 (dotted), 10 (solid), 
25 (large dash). In this case the spin independent cross section can fall 
below the very small spin dependent cross section where such cancellations 
do not occur\cite{bednyakov}.

\begin{figure}[htb]
\centerline{ \DESepsf(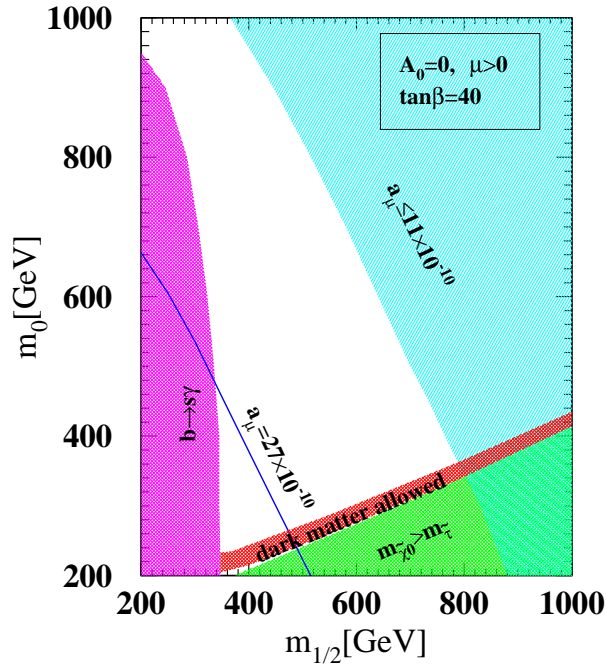 width 7 cm) }
\caption {\label{fig4} Allowed parameter space in the $m_0$ -$m_{1/2}$ plane 
for $\tan\beta = 40$, $A_0 
= 0$, $\mu >0$. The upper (blue) region would be eliminated if $a_\mu < 
11\times10^{-10}$, and the line for $a_\mu = 27\times10^{-10}$ is shown.}
\end{figure}

There exist now two other experiments that might restrict the parameter 
space even more: The BNL E821 \cite{BNL} $g_{\mu} - 2$ experiment measuring the muon 
magnetic moment, and the decay $B_s\rightarrow \mu^+ \mu^-$ which may be observable at the 
Tevatron RunIIB (or possibly in B-factories), and we turn to consider these 
next.

\begin{figure}[htb]
\centerline{ \DESepsf(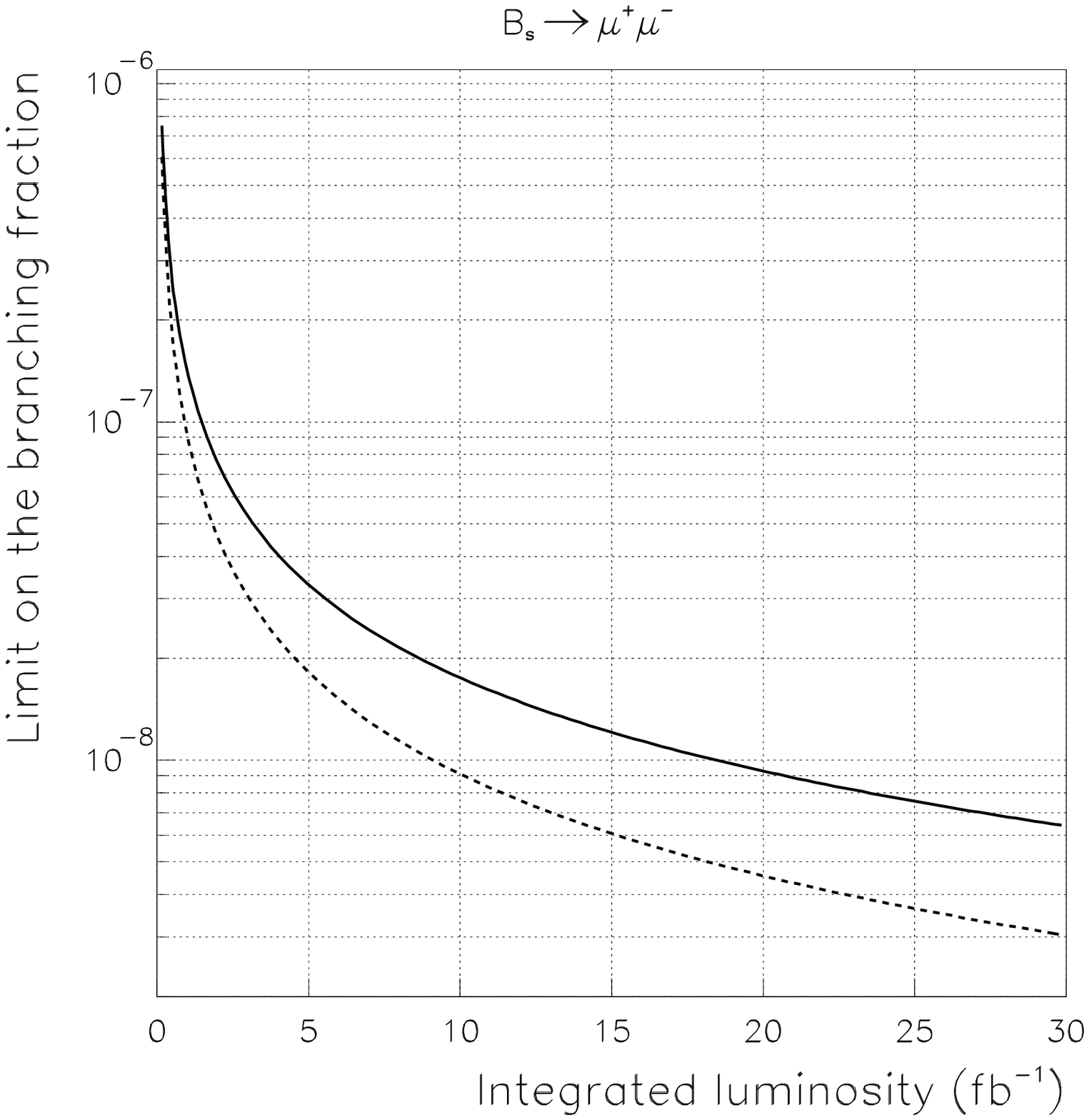 width 7 cm) }
\caption {\label{fig5}  Illustrated 95\% C.L. limits on the branching ratio for
	$B_s\rightarrow \mu^+ \mu^- $ at CDF in Run II as a function of
	integrated luminosity.
	Solid (Case A) and dashed (Case B) 
	curves are based on different assumptions
	on the signal selection efficiency and the background
	rejection power[30].}
\end{figure}

\section{ Muon magnetic Moment Anomaly}

The BNL E821 experiment has measured the muon magnetic moment with 
exceedingly high accuracy. When compared with the calculations of $a_\mu$ 
expected from the Standard Model (with corrected sign of the hadronic 
scattering of light by light contribution \cite{knecht}) there remains a small 
discrepancy:
\begin{equation}\delta a_\mu = 27(16)\times10^{-10} 
\end{equation}
which is a 1.6 $\sigma$ effect. While this is not enough to presume that a real 
effect has been discovered, it is still interesting to examine the effects 
such an anomaly would have for two reasons: first the errors will shortly 
be significantly reduced, and second SUGRA models imply the existence of 
such an anomaly of just this size.

Much of the uncertainty in the calculation of $a_\mu$ comes from the hadronic 
part, $a_\mu$(had). However, it is possible to express this by a dispersion 
relation in terms of integrals over the experimental cross section 
$\sigma_{e^+e^-}\rightarrow had$. Recent experiments from from CMD-2, VEPP-2M and Beijing 
\cite{cmd,achasov,bes} have re-measured these cross sections with greatly improved 
accuracy. Further, the BNL E821 experiment has about six times more data 
which they are currently analyzing which should greatly reduce the 
statistical part of the error. Thus one may expect the error in Eq.(5) to 
be reduced by a factor of two or more in the near future. If the central 
value of the anomaly were to remain unchanged, the effect would become more 
statistically significant.

 From the theoretical side, it has been known for some time that mSUGRA 
predicts an important contribution to $a_\mu$ \cite{g-2,g-22}. This is illustrated in 
Fig. 4 which shows the allowed regions in the $m_0$ - $m_{1/2}$ plane and assumes 
that the entire $a_\mu$ anomaly is due to SUSY. The upper right (blue) region 
corresponds to $a_\mu$ less than $1\sigma$ below the current central value, while 
the diagonal line corresponds to the central value, which falls at the 
lower edge of the allowed part of the parameter space.  The current central 
value thus corresponds to a relatively low mass SUSY spectra, easily 
accessible to the LHC. In fact, too big a value of delta $a_\mu$ (i. e. delta 
$a_\mu\stackrel{>}{\sim}40\times10^{-10})$ would be sufficient to exclude mSUGRA
\cite{bdutta2}. On the 
other hand, an $a_\mu$ less than $1 \sigma$ below the current central value would 
imply a heavy SUSY mass spectrum. Further, if $\delta a_\mu$ is positive, then 
$\mu > 0$ \cite{nano,chat}, and this would eliminate the very low dark matter detection 
cross sections shown in Fig.3. Thus the final value of $\delta a_\mu$ will play 
an important role in deciphering the SUSY parameter space.

\begin{figure}[htb]
\centerline{ \DESepsf(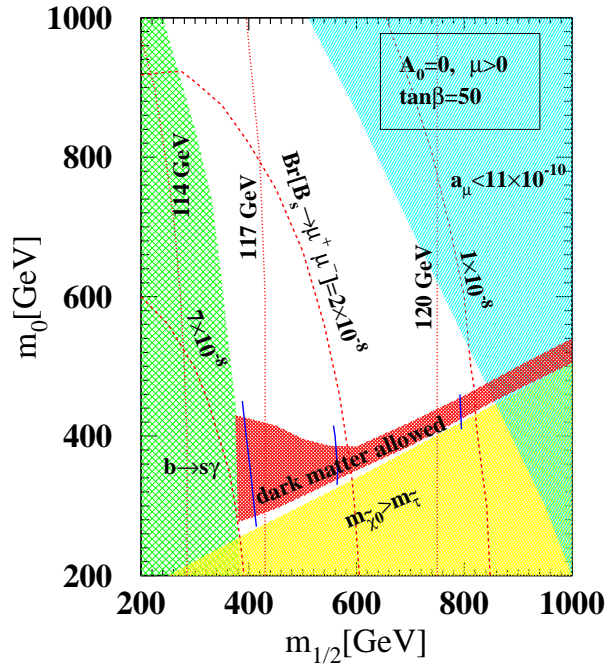 width 7 cm) }
\caption {\label{fig6}  Branching ratio for $B_s\rightarrow \mu^+ \mu^- $ (three dashed lines from left
to right: 
$7\times 10^{-8}$, $2\times 10^{-8}$, $1\times 10^{-8}$) for $\tan\beta$ = 50 in the
$m_0$-$m_{1/2}$ plane. Other mSUGRA parameters are fixed to be 
$A_0 = 0$ and $\mu > 0$.
 The three short solid lines indicate the $\sigma_{\tilde\chi^0_1-p}$ values 
(from left:  0.05 $\times 10^{-6}$ pb,  0.004 $\times 10^{-6}$ pb, 0.002 $\times
10^{-6}$ pb).  The vertical dotted lines label Higgs masses[30].}
\end{figure}

\begin{figure}[htb]
\centerline{ \DESepsf(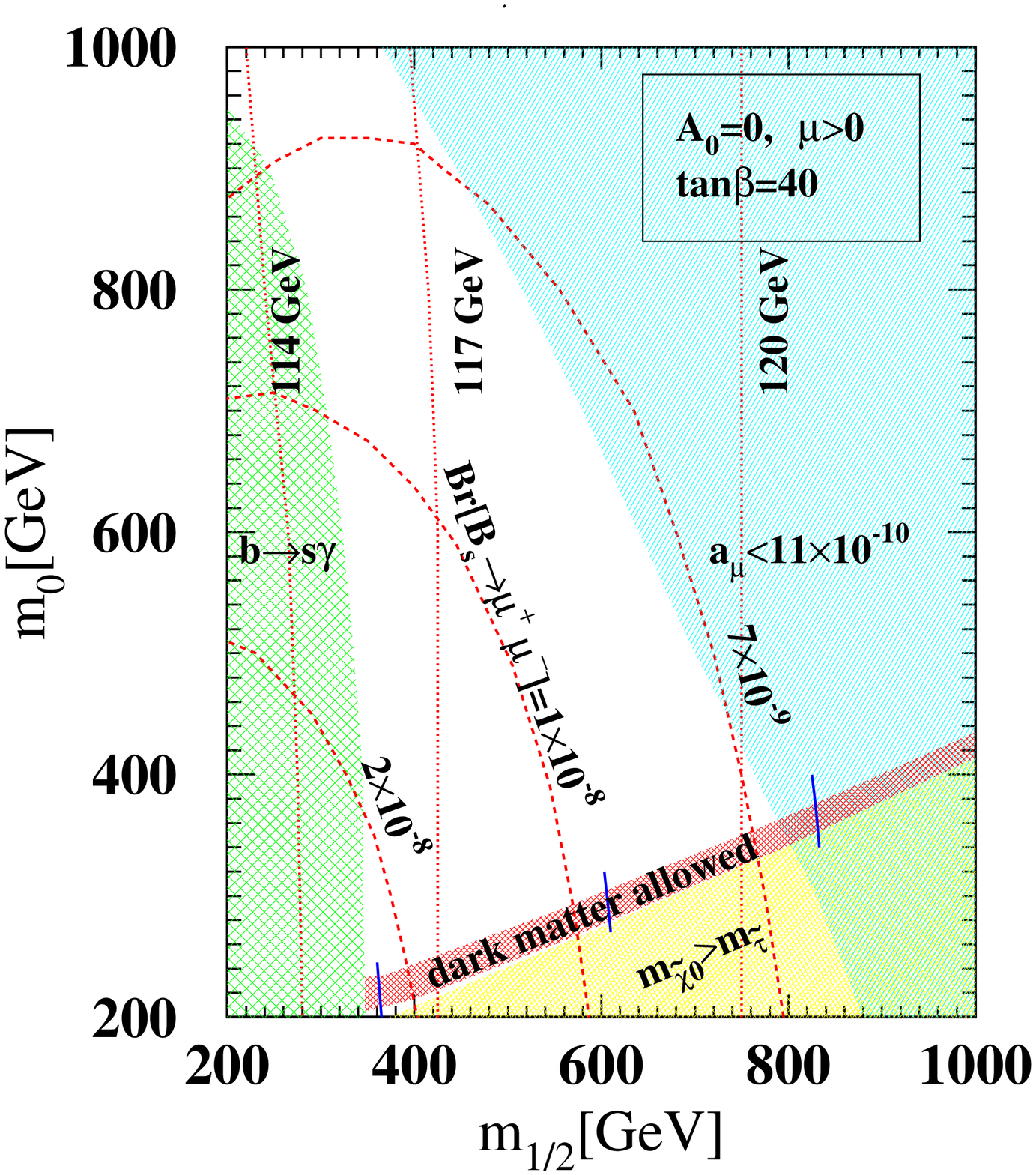 width 7 cm) }
\caption {\label{fig7}  Branching ratio for $B_s\rightarrow \mu^+ \mu^- $  (three dashed lines from left
to right: $1.9\times 10^{-8}$, $1\times 10^{-8}$, 
$0.7\times 10^{-8}$) at $\tan\beta$ = 40 in the $m_0$-$m_{1/2}$ plane. Other
mSUGRA parameters are fixed to be
$A_0 = 0$ and $\mu > 0$.
 The three short solid lines indicate the $\sigma_{\tilde\chi^0_1-p}$ values 
(from left: 0.03 $\times 10^{-6}$ pb,  0.002 $\times 10^{-6}$ pb, 0.001 $\times
10^{-6}$ pb).  The vertical dotted lines label Higgs masses[30].}
\end{figure}

 \begin{figure}[htb]
\centerline{ \DESepsf(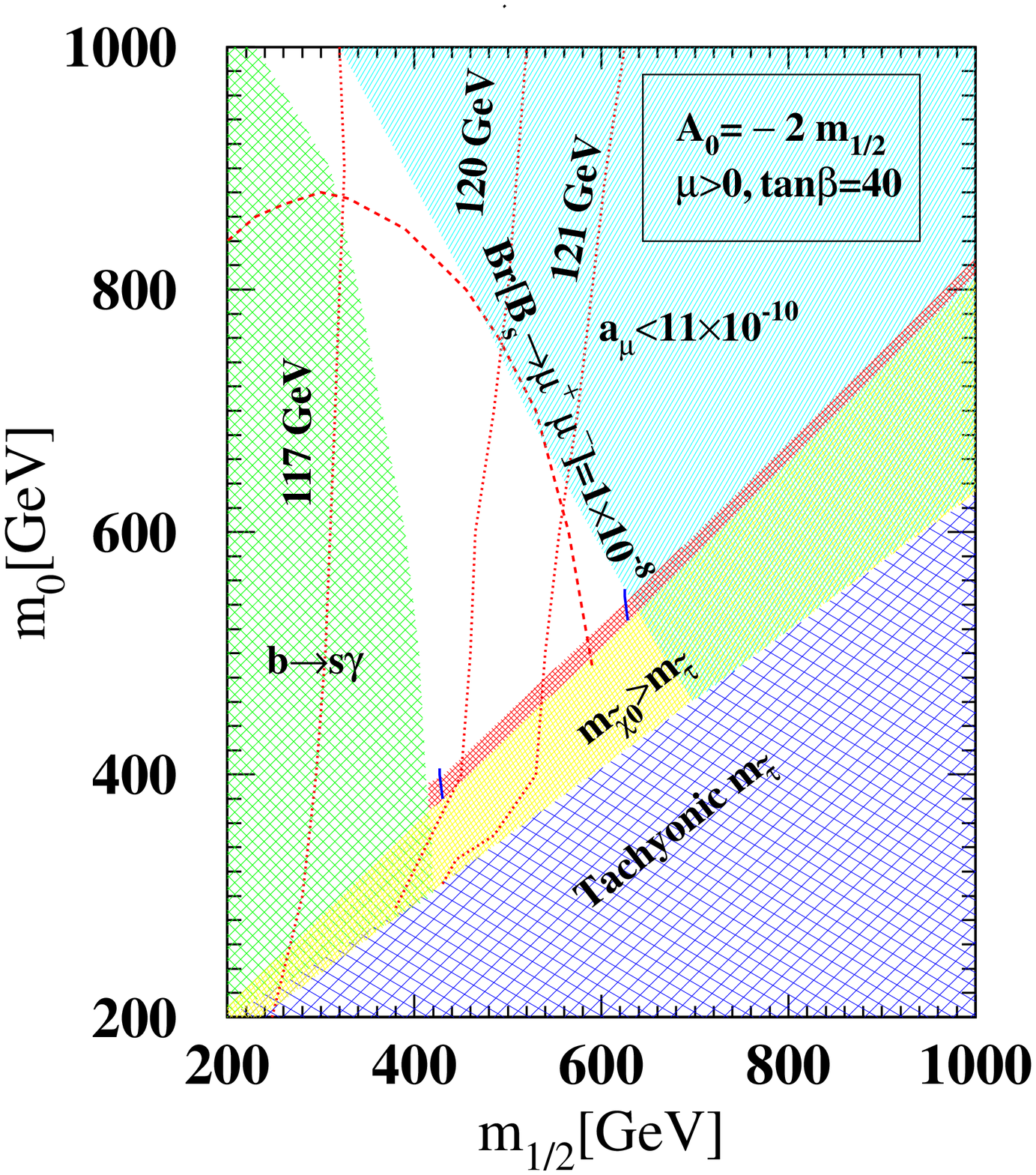 width 7 cm) }
\caption {\label{fig8} Branching ratio for $B_s\rightarrow \mu^+ \mu^- $  at
$\tan\beta$ = 40 in the
$m_0$-$m_{1/2}$ plane for $A_0=-2m_{1/2}$
 and $\mu > 0$. The two short solid lines indicate the
$\sigma_{\tilde\chi^0_1-p}$ values (from left: 
 0.005$\times 10^{-6}$ pb, 0.001 $\times 10^{-6}$ pb). 
 The vertical dotted lines label Higgs masses[30].}
\end{figure}

\section{$B_s\rightarrow \mu^+ \mu^-$}

The $B_s\rightarrow \mu^+ \mu^-$ decay offers an additional window for investigating the 
mSUGRA parameter space. This process has been examined within the MSSM 
framework \cite{babu,bobeth} and more recently using mSUGRA \cite{dedes}. We consider here 
predictions for this decay for mSUGRA, but include all the current 
experimental constraints listed in Sec. 3  (which are necessary to see what 
predictions occur\cite{bdutta3}). The $B_s\rightarrow \mu^+ \mu^-$ decay is of interest since the 
Standard Model prediction for the branching ratio is quite small 
\cite{bobeth}($Br[B_s\rightarrow \mu^+ \mu^-] = (3.1 \pm1.4)\times10^{-9}$), while the SUSY contribution can become 
quite large for large $\tan\beta$. This is because the leading diagrams grow as 
$(\tan\beta)^3$ and hence the branching ratio as $(\tan\beta)^6$. What further 
makes this decay interesting is that it is possible to find a set of cuts 
so that CDF (and probably also DO) may be able to observe it in Run2B (with 
15 $\rm fb^{-1}$ data). Fig. 5 shows the CDF limit on the branching fraction as a 
function of the luminosity\cite{bdutta3}. (The solid curve is a conservative 
estimate, and the dotted curve a more optimistic possibility.) One sees 
that  CDF will be sensitive to a branching ratio of $Br > 1.2\times10^{-8}$ (and 
the combined CDF and D0 data to $Br >  6.5\times 10^{-9}$). mSUGRA analysis then 
shows that CDF would be sensitive to this decay for 
$\tan\beta \stackrel{>}{\sim}30$.

\begin{figure}[htb]
\centerline{ \DESepsf(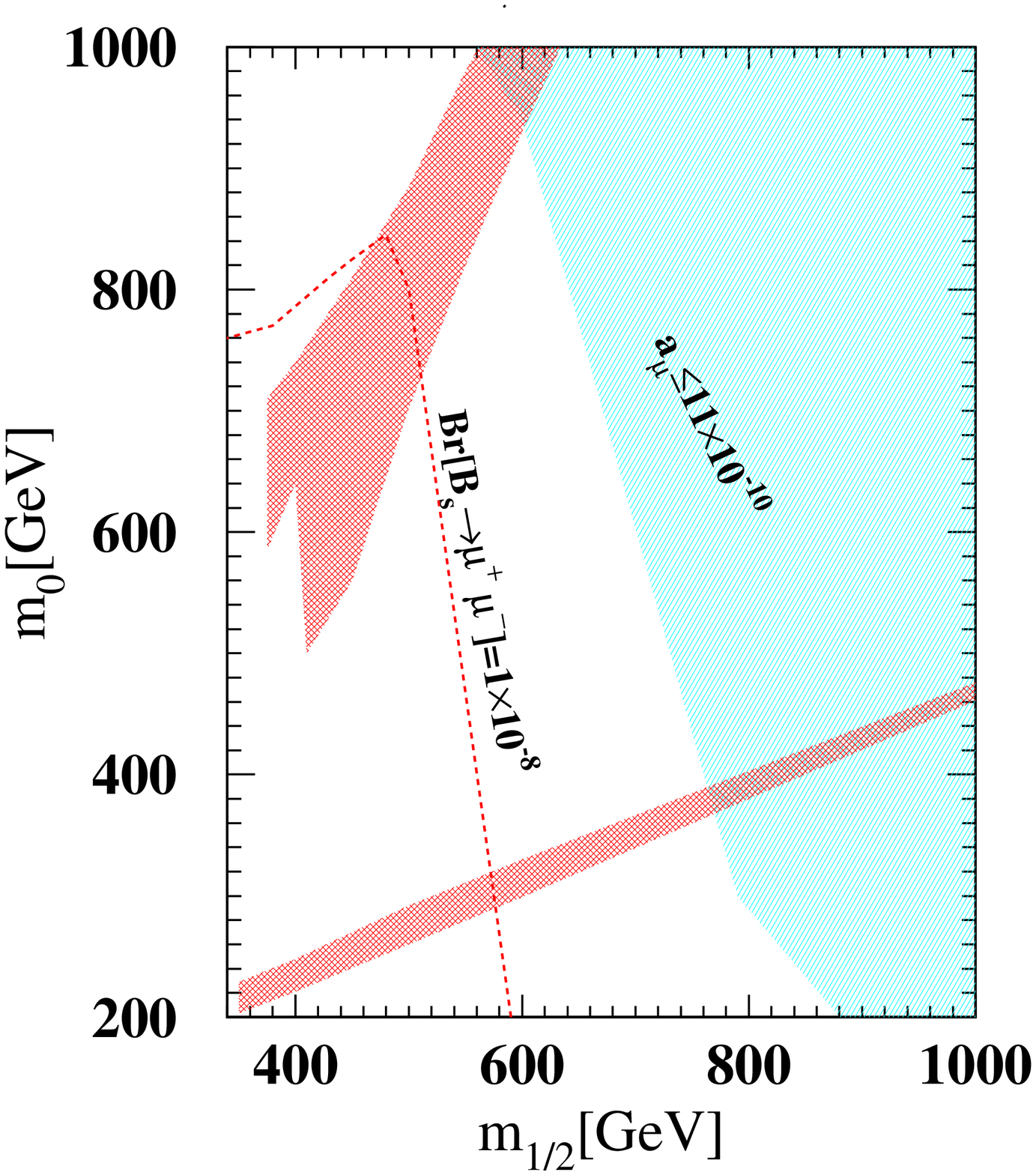 width 7 cm) }
\caption {\label{fig9}Allowed regions in the $m_0$ - $m_{1/2}$ plane for $\delta_2 = 1$, 
$\tan\beta = 
40$, $A_0 = 0$, $\mu >0$. The lower band is the ususal 
$\tilde\tau_1$-$\tilde\chi^0_1$ co-annihilation 
channel, and the upper region is the new  $Z$ boson s-channel annihilation 
region. }
\end{figure}

\begin{figure}[htb]
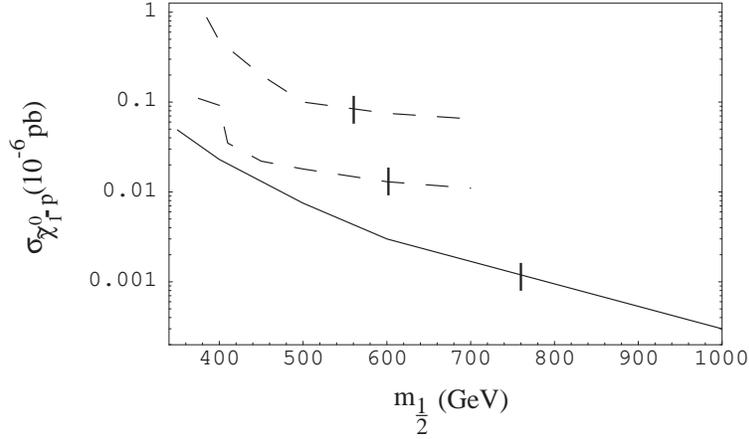

\centerline{ \DESepsf(adhs4.epsf  width 10 cm) }
\caption {\label{fig10} $\sigma_{\tilde{\chi}_1^0-p}$ as a function of $m_{1/2}$
($m_{\tilde{\chi}_1^0} \stackrel{\sim}{=} 0.4 m_{1/2}$) for $\tan\beta = 40$, 
$\mu >0$, $m_h > 114$ GeV, $A_0 = m_{1/2}$ for $\delta_2 = 1$. The lower curve
is for the $\tilde{\tau}_1-\tilde{\chi}_1^0$ co-annihilation channel, and the
dashed band is for the $Z$ s-channel 
annihilation allowed by non-universal soft breaking. The curves terminate 
at low $m_{1/2}$ due to the $b \rightarrow s\gamma$ constraint. The vertical
lines show the 
termination at high $m_{1/2}$ for $\delta a_{\mu}>11\times 10^{-10}$[24].}
\end{figure}

We can now examine the effects of the combination of all data. Figs. 6 and 
7 \cite{bdutta3}[show the parameter space for $\tan\beta = 50$ and $\tan\beta =40$ 
respectively for $A_0$ = 0 and $\mu > 0$.  One sees that there is an upper bound 
on the $Br[B_s\rightarrow \mu^+ \mu^-]$ to be consistent with mSUGRA, and a branching 
ratio $> 7(14)\times 10^{-8}$ would be able to exclude mSUGRA for $\tan\beta < 
50(55)$. As can be seen from Fig. 5, such a branching ratio could be seen 
with only 2 $\rm fb^{-1}$. More generally, Fig. 6 shows that the entire parameter 
space can be covered for $\tan\beta = 50$, $A_0 = 0$ if 
$a_\mu > 11\times10^{-10}$, and from Fig. 7 also for $\tan\beta = 40$ using the
combined CDF and D0 data. The effect of varying  $A_0$ 
is shown in Fig. 8 where $A_0 = -2m_{1/2}$, and $\tan\beta = 40$. Here again the full 
parameter space can be covered.

One sees from the above graphs that future measurements should be able to 
determine the basic parameters of the mSUGRA model. Thus   when the lines 
for $\delta a_\mu$, $Br[B_s \rightarrow\mu^+ \mu^-]$ and $m_h$ intersect the allowed dark matter band at 
a single ``point" for a given $A_0$, this will determine $m_0$ (approximately), 
$m_{1/2}$ and $\tan\beta$. (It may be better to use the gluino mass in place 
of $m_{h}$ since the parameter space is very sensitive to $m_h$.)  Predictions of 
the SUSY spectrum and dark matter detection rates would then follow.

\section{Non-Universal Models}

One can generalize mSUGRA by allowing non-universal soft breaking at $M_G$ in 
the third generation of squarks and sleptons and also in the Higgs sector. 
If universality of the gaugino masses is maintained, then stau-neutralino 
co-annihilation will still play an important role. However, new effects can 
occur, since the non-universality effects the size of the $\mu$ parameter. The 
$\mu$ parameter governs the Higgsino content of the $\tilde\chi^0_1$ and as $\mu^2$ 
decreases (increases), the Higgsino content increases (decreases). Since 
$\sigma_{\tilde\chi^0_1-p}$ depends on the interference between the Higgsino and gaugino 
parts of $\tilde\chi^0_1$,  $\sigma_{\tilde\chi^0_1-p}$ will correspondingly increase (decrease). A second 
effect also occurs. As the Higgsino content of the $\tilde\chi^0_1$ increases, then the 
$\tilde\chi^0_1$ - $\tilde\chi^0_1$ - $Z$ coupling is strengthened, allowing a new annihilation 
channel to become important (in addition to the $\tilde\tau_1$-$\tilde\chi^0_1$
 co-annihilation 
channel). As a simple example we consider the case where at $M_G$ one chooses 
$m_{H_2}^2 = m_0^2 (1 + \delta_2)$, and all other soft breaking masses universal. 
Fig. 9 shows the allowed region in the $m_0$ - $m_{1/2}$ plane for $\tan\beta = 40$, 
$\delta_2 = 1$, $A_0 = 0$, $\mu >0$. One sees the usual narrow stau-neutralino 
co-annihilation band a relatively low $m_0$, but in addition there is a 
higher $m_0$ (and low $m_{1/2}$) region satisfying all constraints due to the new 
$Z$-channel annihilation process. Fig. 10 shows  $\sigma_{\tilde\chi^0_1-p}$ as a function of 
$m_{1/2}$ for $\tan\beta = 40$, $A_0 = m_{1/2}$, $\mu >0$. The Z-channel corridor now reaches 
up to the DAMA data region \cite{belli} for low $m_{1/2}$ ($m_{\tilde\chi^0_1} \simeq 0.4m_{1/2}$), and so if 
the DAMA results are confirmed, it would point to a non-universality of 
this type.

\section{Conclusions}

We have considered here SUGRA models with R parity invariance and grand 
unification at $M_G \simeq 2\times10^{16}$ GeV. Current data has begun to constraint 
these models significantly. For mSUGRA (and many non-universal models), 
accelerator data ($m_h$ and $b \rightarrow s\gamma$) place lower bounds on 
$m_{1/2}$ such that 
$m_{1/2} \stackrel{>}{\sim} 300$GeV (or $m_{\tilde\chi^0_1} \stackrel{>}{\sim} 120$GeV), while astronomical data on the amount of 
dark matter narrowly determines m0 in terms of $m_{1/2}$ 
(for each $A_0$ and $\tan\beta$).

Thus additional new data could begin to fix the parameters of mSUGRA 
parameters, and so determine the mass spectrum expected at the LHC, dark 
matter detection rates, etc. In particular, the muon magnetic moment 
anomaly $a_\mu$ and the branching ratio $Br[B_s \rightarrow\mu \mu]$ could combine with 
$m_h$ or a measurement of $m_{\tilde g}$  to fix the parameters completely. A more 
accurate determination of $a_\mu$ should be available shortly, and we have 
seen that the Tevatron RunII should be sensitive to the $B_s$ decay for $Br > 
1.2(0.65)\times 10^{-8}$ for the CDF (or combined CDF and D0) data, and this would 
cover a large part of the parameter space for $\tan\beta > 30$. Non-universal 
models in the Higgs and third generation soft breaking masses offer the 
possibility of a new $Z$ boson s-channel annihilation in the early universe. 
Such possibilities can significantly increase the neutralino-proton 
detection cross section up the DAMA region of $10^{-6}$ pb.

\section{Acknowledgement} This work was supported in part by National Science
Foundation Grant PHY-0101015.

\end{document}